# Borophosphene as a promising Dirac anode with large capacity and high-rate capability for sodium-ion batteries

Yang Zhang[*], Er-Hu Zhang, Ming-Gang Xia, Sheng-Li Zhang

*Ministry of Education Key Laboratory for Nonequilibrium Synthesis and Modulation of Condensed Matter, Department of Applied Physics, School of Science, Xi'an Jiaotong University, Xi'an 710049, China*

**ABSTRACT**

Sodium-ion batteries (SIBs) have attracted a great deal of attention as potential low-cost energy storage alternatives to Lithium-ion batteries (LIBs) due to the intrinsic safety and great abundance of sodium on Earth. For developing competitive SIBs, highly efficient anode materials with large capacity and rapid ion diffusion are indispensable. In this study, a two-dimensional (2D) Dirac monolayer, that is borophosphene, is proposed to be a promising anode material for high performance SIBs on the basis of density functional theory calculations. The performances of Na adsorption and diffusion, maximum specific capacity, open circuit voltage, cyclical stability and electronic properties combined with Bader charge analysis are explored. It is found that the borophosphene can spontaneously adsorb Na atom with binding energy of -0.838 eV. A low diffusion energy barrier of 0.221 eV suggests rapid ion conductivity. More intriguingly, a maximum specific capacity of 1282 mAh/g can be achieved in borophosphene, which is one of the largest values reported in 2D anode materials for SIBs. A low average voltage of 0.367 V is estimated, implying a suitable voltage of the anode material. Metallic properties, tiny surface expansion, and good kinetic stability of sodiated borophosphene give rise to high electrical conductivity and favorable cyclability. These advantages above suggest the borophosphene can be used as a Dirac anode material for SIBs with excellent performances of large specific capacity, high-rate capability, and favorable cyclability.

## 1. INTRODUCTION

With the rapid development of microelectronics industry and electrification of transportation, the demand for rechargeable batteries with high energy and power densities are highly desirable.[1-3] Lithium-ion batteries (LIBs), one of the most prevailing batteries, have been widely used in portable electronic products and fully electric vehicles due to their great advantages of high energy density, long cycle lifespan and non-pollution to the environment.[4] However, the limited storage and uneven distribution of Li on Earth have greatly blocked the large-scale application of LIBs. Sodium-ion batteries (SIBs), owning to the great abundance of Na storage, are promising low-cost energy storage alternatives to LIBs. Moreover, the intrinsic safety of SIBs is much better than that of current LIBs.[5] Since Na and Li atoms are the same-group elements in the periodic table and share common physical-chemical properties as alkali metals. Thus, the mature technologies developed in LIBs can be transferred to SIBs. However, due to the large atomic radius of Na, well-developed anode materials used in LIBs are unsuitable for SIBs, resulting in low specific capacity and energy density, and poor cycle life.[6,7] Therefore, a promising strategy is to develop new anode materials of SIBs which behave high electrical conductivity for rapid electron transport and large surface areas with a shallow energy barrier for fast Na diffusion.

Two-dimensional (2D) nanostructures, including borophene,[8-10] phosphorene,[11,12] metal oxides,[13,14] metal sulfides,[14-17] and transition-metal carbides (MXenes),[18-20] have been proposed as potential anode materials for SIBs owning to their large surfaces that can provide lots of adsorption sites for Na atoms. Unfortunately, many of them are semiconductors with low electronic conductivity, and most of them exhibit low specific capacity. Due to the ultrahigh electronic conductivity and large specific area, 2D Dirac nanosheets have stimulated great interest as binder-free anodes. The most prominent 2D Dirac anode is graphene; however, the pristine graphene is inactive toward metal ions. By inducing lattice defects or extra edges in graphene, the electrochemical ion storage can be highly improved.[21] On the contrary, other 2D Dirac materials, including silicene, germanene and stanene, exhibit better affinity to metal ions;[22-26] however, their self-weight is higher than graphene, resulting in a low specific capacity. For examples, low Na capacities of 369 and 226 mAh/g are predicted for germanene and stanene, respectively.[23] By inducing nanohole defects in stanene to reduce its self-weight, the theoretical capacity can be greatly improved to 363 mAh/g.[26] For practical applications, these specific capacities are hard to satisfy the development of science and technology.

Very recently, Dirac monolayers composed of light elements are proposed to be promising as anode materials with high specific capacity for SIBs. Borophane, a hydrogenated borophene, is a gapless Dirac material. It can behave a high capacity of 504 mAh/g, over 2 times higher than that of stanene.[27] Using a global structure search method combined with first-principles calculations, a new 2D anisotropic Dirac cone material, $B_2S$ monolayer, has been identified to be stable.[28] Owning to its light weight, the $B_2S$ monolayer can exhibit a theoretical capacity of 998 mAh/g, as well as a low diffusion energy barrier of 0.19 eV,[29] suggesting a lightweight 2D Dirac anode material with high capacity and desirable rate performance. In our previous study, a new 2D Dirac monolayer of borophosphene has been proposed to be stable mechanically, thermally and dynamically, which is composed of light elements of B and P.[30] Interestingly, the Dirac cone of borophosphene is robust, independent of in-plane biaxial and uniaxial strains, and can also be observed in its one-dimensional nanoribbons and nanotubes, exhibiting high electronic conductivity with the Fermi velocity of $10^5$ m/s. These advantages imply that the borophosphene might be expected as a promising electrode material for SIBs. To explore it, a detailed study on Na adsorption and diffusion process on borophosphene is indispensible.

In this study, we will employ first-principles study to address the potential of borophosphene as a Dirac anode material for SIBs. Various binding sites and different pathways are considered for Na adsorption and diffusion. Bader charge analysis is used to characterize the interaction of ion bonds between Na adatom and borophosphene. Intriguingly, a high specific capacity of 1282 mAh/g can be achieved, which is one of the largest values for existing 2D anode materials. A low diffusion energy barrier of 0.221 eV suggests fast ion diffusion. Open circuit voltage and electronic conductivity are explored during the charging process. Using *ab* initio molecular dynamic (AIMD) simulations, the cyclical stability of fully sodiated borophosphene is examined from the aspects of dynamic stability and surface expansion.

## 2. COMPUTATIONAL METHODS

All calculations were performed using Vienna *ab initio* simulation package (VASP) based on the spin-polarized density functional theory (DFT) within the projector augmented wave method.[31-34] The generalized gradient approximation (GGA) with the functional of Perdew-Burke-Ernzerhof (PBE) was employed to describe the electron exchange-correlation interaction.[35,36] The energy cutoff of the plane-wave basis expansion and the convergence of total energy were set to 400 and $10^{-5}$ eV. A supercell of borophosphene containing $3 \times 2$ primitive cells was adopted as shown in Figure 1a, which corresponds to the cell sizes of 9.66 and 11.13 Å perpendicular and parallel to the direction of B-B or P-P dimer. To eliminate the interactions between neighbor layers, a vacuum space of 15 Å was applied along the normal direction of borophosphene sheet. Thus, the Brillouin zone was sampled using a $9 \times 7 \times 1$ *k*-point mesh generated by the Monkhorst-Pack scheme. For structural optimizations, the Hellmann-Feynman forces using conjugate gradient algorithm were calculated within a force convergence of 0.01 eV/Å.[37] In order to accurately evaluate the interactions between Na adatoms and borophosphene, van der Waals (vdW) corrections with DFT-D2 method of Grimme were considered.[38]

During the charging process, Na ions migrate from the cathode electrode, finally insert into the anode electrode and load on the surface. Thereby, the binding strength of the Na atoms on the anode surfaces plays a key role for improving the performance of SIBs. To evaluate this strength, binding energy is calculated as the following equation:

$$E_b = [E_{BP+Na_n} - (E_{BP} + nE_{Na})] / n, \qquad (1)$$

where $E_{BP+Na_n}$ and $E_{BP}$ are the total energies of the borophosphene with and without Na adsorption. $E_{Na}$ is the binding energy per Na atom in the *bcc* bulk. $n$ is the number of Na atoms. According to this definition, a negative $E_b$ can be obtained. A smaller value represents more favorable binding stability.

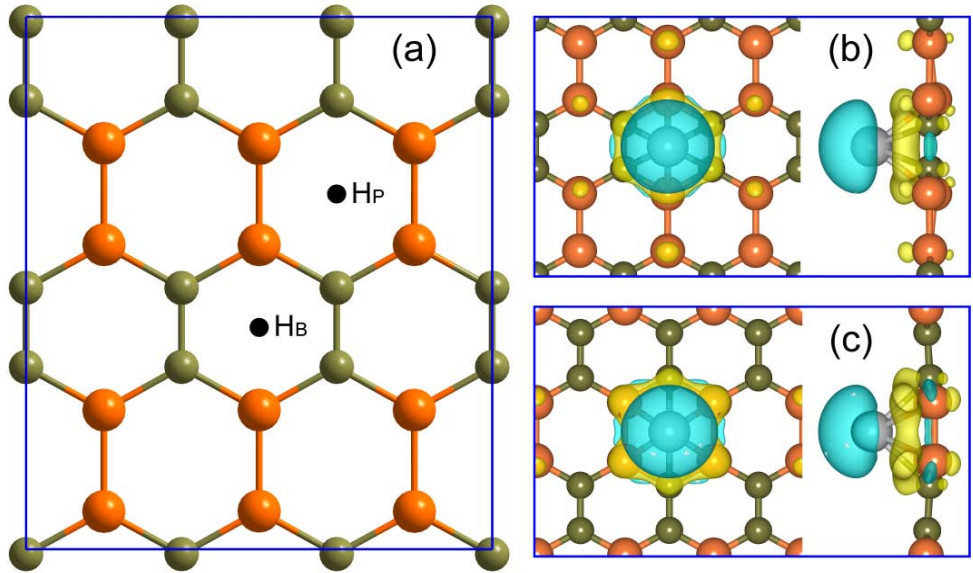

**Figure 1.** (a) Schematic illustration of single Na adsorbing on borophosphene surfaces and corresponding differences of charge density for single Na adsorptions at (b) $H_B$ and (c) $H_P$ sites. Yellow and turquoise isosurfaces represent electron accumulation ($\Delta\rho>0$) and depletion ($\Delta\rho<0$), and the isosurface level is set to 0.004 $e/\text{Å}^3$. Dark yellow, orange and gray balls denote B, P and Na atoms, respectively.

## 3. RESULTS AND DISCUSSION

The optimized geometrical structure of borophosphene is shown in Figure 1a. Freestanding borophosphene is a planar and atomic nanosheet composed of B-B and P-P dimers, exhibiting orthogonal symmetry with two B and two P atoms in a unit cell, which has been proposed by using our in-house code to implement the genetic algorithms method.[39] The calculated lattice constants are 3.22 and 5.57 Å, which are very close to those of hexagonal boron-phosphide monolayer.[40-42] Such a planar sheet has been demonstrated to be dynamically, thermally and mechanically stable in our previous study.[30] Electronic band structure and density of state of the borophosphene are calculated. The results are shown in Figure S1. It can be seen that, owning to the occupations of only the $p_z$ orbitals from both the B and P atoms, a zero-gap Dirac cone with $\pi$ and $\pi^*$ bonds is observed at the $k$-point of D = $\pi/a$·(0.75, 0, 0). The Fermi velocity is estimated to $6.51 \times 10^5$ m/s, which is in the same order of magnitude with that of graphene ($8.2 \times 10^5$ m/s),[43] suggesting excellent conductivity of borophosphene.

**Table 1**. Properties of Single Na Atom Adsorbing on Borophosphene: Average Bond Lengths of Na-B and Na-P $l$ (Å), the Shortest Distance between Na Adatom and Borophosphene $d$ (Å), Binding Energy $E_b$ (eV) and Amount of Bader Charge Transferred from Na Adatom to Borophosphene $\Delta$Q ($e$).

| binding site | $l_{\text{Na-B}}$ | $l_{\text{Na-P}}$ | $d$ | $E_b$ | $\Delta$Q |
|---|---|---|---|---|---|
| $H_B$ | 2.798 | 2.897 | 2.040 | -0.838 | 0.851 |
| $H_P$ | 2.847 | 2.968 | 1.920 | -0.707 | 0.858 |

As an efficient electrode material for SIBs, spontaneous adsorption and rapid diffusion of Na atoms are fundamental factors, which can largely affect the performances of charging/discharging and circuit rate. For exploring those properties, seven possible adsorption sites are considered (see Figure S2). For each case, the initial position of Na adatom is set to above the borophosphene surface with a height of 4 Å or so, beyond the common distances of chemical bonds. After fully geometrical optimization, Na adatom can automatically migrate to the borophosphene surface and then adsorb on it, which means an exothermic process with no energy barrier. As shown in Figure 1a and S2, only two adsorption configurations are stabilized, namely $H_B$ and $H_P$, where Na adatoms are located at the centers of six-membered $B_4P_2$ and $B_2P_4$ rings, respectively. The structural parameters and binding energies are summarized in Table 1. Because of smaller bond lengths of Na-B and Na-P, the $H_B$ site is more stable than the $H_P$ for Na adsorption, exhibiting that Na atom prefers to bind with B atoms. Such a combining hobby can also be deduced from the slight distortion of P atoms out of the borophosphene plane. In view of the charge density difference (see Figure 1b and 1c) or the Bader charge analysis (see Table 1), strong ionic Na-B and Na-P bonds can be formed between the borophosphene and Na adatom. Obvious electron depletion occurs around the Na adatom, while electron accumulation merges on the borophosphene electrode, which is mainly localized on the six-membered rings beneath the adsorbed ions. The amount of Bader charge transferred from the Na adatom to borophosphene is about 0.85 e for both the $H_B$ and $H_P$ sites. Thereby, the calculated binding energies are -0.838 and -0.707 eV for the $H_B$ and $H_P$ sites, which are much lower than those of other Dirac anode materials, such as graphene (0.2~0.4 eV)[44,45] and $B_2S$ (-0.29 eV).[29] Generally, a lower binding energy for a certain electrode suggests stronger binding strength. But, it might result in a larger energy barrier for ion diffusion.

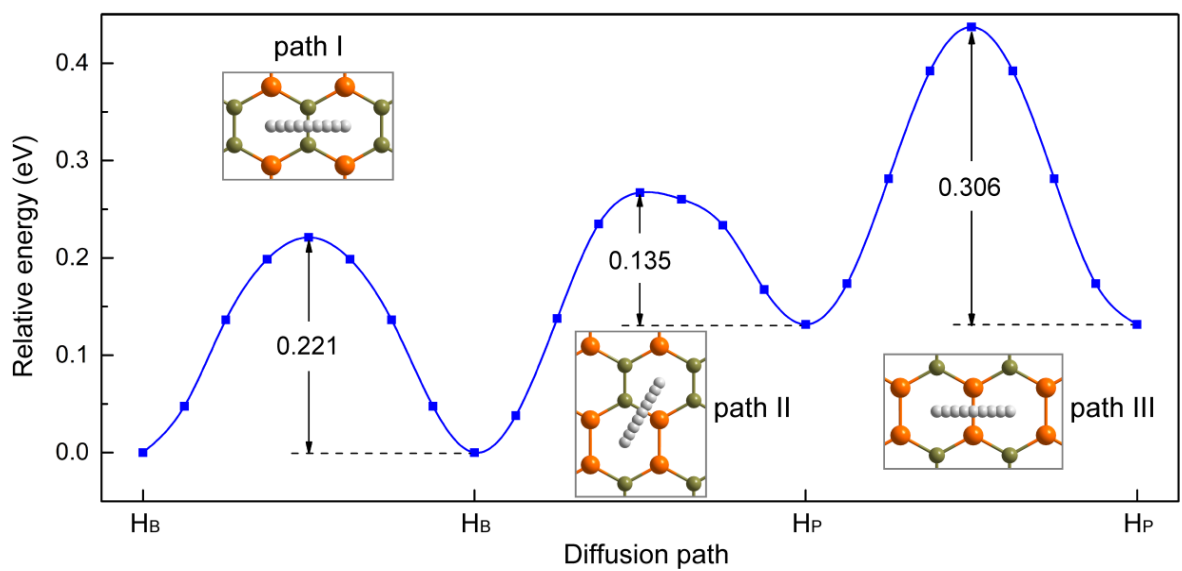

**Figure 2.** Energy barriers of Na atom migrating along various diffusion pathways: path I ($H_B \rightarrow H_B$), path II ($H_B \rightarrow H_P$) and path III ($H_P \rightarrow H_P$). Inserts display the corresponding diffusion pathways.

For exploring the diffusivity of Na atom, three pathways, that is, path I ($H_B \rightarrow H_B$), path II ($H_B \rightarrow H_P$) and path III ($H_P \rightarrow H_P$) are considered, as displayed in Figure 2. The path I and path III are perpendicular to the directions of B-B and P-P dimers. Diffusing along these two paths requires skipping over the dimers directly. For the path I, the diffusion energy barrier is 0.221 eV. Whereas, a little large barrier of 0.306 eV is obtained for the path III. These energy barriers suggest that the diffusion along the $H_P \rightarrow H_P$ is slightly harder than along the $H_B \rightarrow H_B$, which is partially because Na atoms prefer to bind with B atoms. But, for diffusing along the path II, it is much easier. The energy barrier is 0.135 eV for diffusing from $H_P$ to $H_B$ site. Thereby, for Na diffusion from the $H_P$ site, a feasible pathway is $H_P \rightarrow H_B \rightarrow H_B$ with a small energy barrier of 0.221 eV. In fact, the diffusion of Na atom might occur parallel to the directions of B-B and P-P dimers. In view of this point, two possible pathways, namely path A and path B are calculated, as shown in Figure S3. The diffusion barrier is relatively high (0.437 eV). Since the adsorption site at the bridge of B-P bond (denoted as site 5 in Figure S2) is a saddle point, a probable diffusion path along the dimer direction is $H_B \rightarrow H_P \rightarrow H_B$, like a "zigzag" route. The corresponding energy barrier equals to the difference of binding energy between the site 5 and $H_B$, that is, 0.267 eV. Thus, it can be concluded that the ability of Na diffusion is less dependent on the directions of borophosphene because of the comparable energy barriers along different pathways (0.221 eV for diffusing perpendicular to the directions of B-B and P-P dimers, and 0.267 eV for parallel to the directions of B-B and P-P dimers). For comparison, the energy barriers of Na diffusion on other 2D anode materials are summarized in Table 2. It can be seen that the energy barrier for Na diffusing on borophosphene is larger than those for $TiC_3$ (0.18 eV),[19] silicene (0.16 eV),[22-24] borophane (0.09 eV)[27] and $PC_3$ (0.05 eV),[46] and is comparable to those for $B_2S$ (0.21 eV)[29] and h-BP (0.217 eV).[47] But, by comparing to other promising electrodes such as $MoS_2$ (0.28 eV)[15] and layered cathodes: $NaCoO_2$ (0.3 eV)[48] and $NaFePO_4F$ (0.29 eV),[49] the diffusion barrier of Na atom on the borophosphene is much smaller, suggesting better diffusion ability.

**Table 2**. Comparison of Diffusion Energy Barrier, Specific Capacity and OCV of Candidate Anode Materials for SIBs.

| anode | energy barrier (eV) | specific capacity (mAh/g) | OCV (V) |
|---|---|---|---|
| borophosphene | 0.221 | 1282 | 0.367 |
| silicene [22-24] | 0.16 | 954 | - |
| germanene [23] | - | 369 | - |
| stanene [23] | - | 226 | - |
| borophane [27] | 0.09 | 504 | 0.03 |
| $B_2S$ [29] | 0.21 | 998 | 0.32 |
| $PC_3$ [46] | 0.05 | 1200 | 0.41 |
| h-BP [47] | 0.217 | 143 | - |

According to the Arrhenius equation, temperature-dependent diffusion rate for Na atom on borophosphene can be evaluated by the following equation $D \propto \exp\left(\dfrac{-E_a}{k_B T}\right)$, where $E_a$ and $k_B$ are diffusion energy barrier and Boltzmann constant, $T$ is the environmental temperature. At room temperature, the diffusion mobility of Na atom on borophosphene along the path I (i.e., $H_B \rightarrow H_B$ shown in Figure 2) is estimated to be about 5.9 times faster than that along the path of $H_B \rightarrow H_P \rightarrow H_B$, implying weak anisotropy of ion diffusivity. For comparison with other electrode materials, the diffusion rate of Na ion on borophosphene is about 9.8, 14.4 and 21.2 times faster than $MoS_2$ and layered cathodes: $NaFePO_4F$ and $NaCoO_2$, respectively, which are promising electrodes for SIBs.

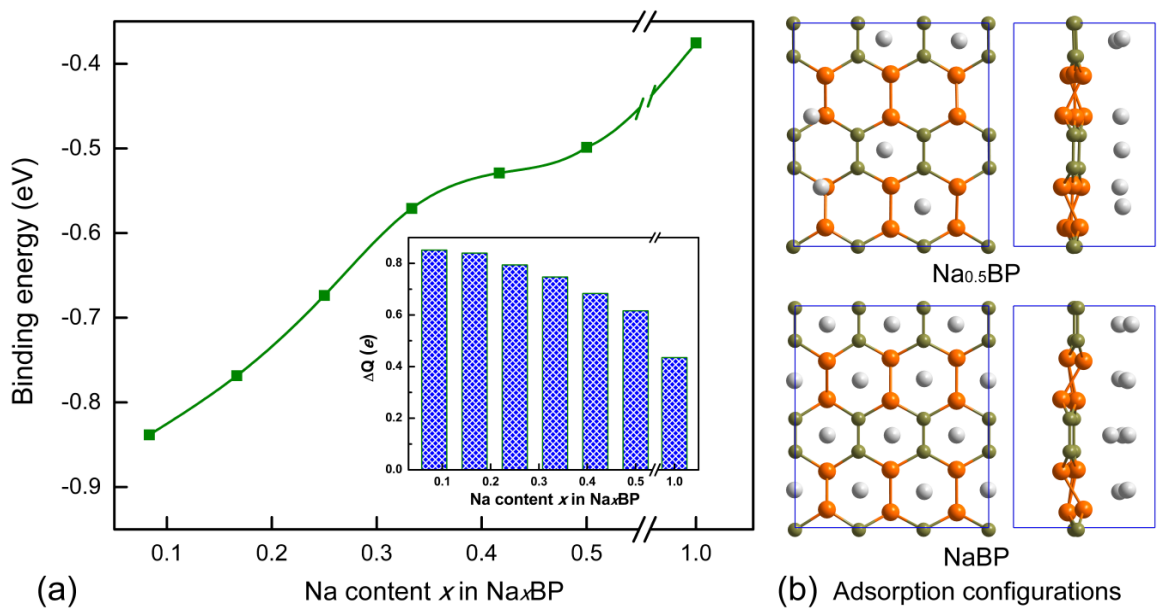

(a)
(b) Adsorption configurations

**Figure 3.** (a) Binding energy as a functional of Na content $x$ in $Na_xBP$ complex and (b) geometrical structures of $Na_{0.5}BP$ and NaBP species. Insert in (a) shows average amount of Bader charge transferred from Na atom to borophosphene against Na content $x$.

To further understand the performance of borophosphene-based SIBs, the structural stability and electronic properties of $Na_xBP$ with respect to Na content $x$ should be explored. By employing a $3 \times 2$ supercell adsorbed with different numbers of Na atoms, a series of concentrations with a stoichiometry of $Na_xBP$ can be constructed ($x = 0.167, 0.25, 0.333, 0.417$ and $0.5$). As shown in Figure S4, for each case, various adsorption configurations are considered. We choose the configuration with the lowest energy as the most stable one. All studied $Na_xBP$ species are predicted to be stable with a low negative binding energy. However, due to the electrostatic repulsive interactions among the neighbor Na cations, the binding energy increases from -0.838 to -0.375 eV as the Na content $x$ increases from 0.083 to 1.0 (see Figure 3a), indicating that the $Na_xBP$ with a higher Na concentration exhibits lower structural stability. Even so, the embedded Na atoms cannot be assembled easily on the borophosphene surface because of the negative binding energy. Only small distortion is observed out of the layer (see Figure 3b), which cannot destroy the geometrical structure of borophosphene. In view of the charge density difference (see Figure S5) and Bader charge analysis (see the insert in Figure 3a), the characteristics of ionic bonds can be observed with obvious amounts of charge transferred from Na adatoms to borophosphene. As the Na content increases, the interactions of ionic bonds gradually weaken. A smaller amount of charge transfer is presented for the Na$_x$BP specie with a higher Na content.

Note, for all studied Na$_x$BP species above, the embedded Na atoms are just located on single side of borophosphene surfaces. As an attempt, we have intercalated 24 Na atoms to double sides of borophosphene surfaces within a large 3 × 2 supercell (i.e., Na$_2$BP) and found that the Na$_2$BP can be stable with a favorable binding energy of -0.267 eV (see Figure S6). Thus, a maximum of two Na atoms per BP formula can be adsorbed on two sides of borophosphene. The theoretical specific capacity of Na atoms on the borophosphene can be evaluated using the following equation:

$$C = \frac{xzF}{M}$$

where $x$ is the concentration of Na atoms; $z = 1$; $F$ is the Faraday constant (26801 mAh/g); $M$ is molecular weight of electrode material. It is found that the specific capacity of borophosphene is 1282 mAh/g, which is one of the highest values in potential 2D anode materials reported for SIBs (see Table 2).

Then, an important issue is to assess the stability of borophosphene during the charging and discharging processes, because the stability of electrode is one of the major factors affecting the performances of batteries. Because of its maximum Na content and relatively high binding energy, we take Na$_2$BP specie as an example to explore the thermal stability by performing *ab initio* molecular dynamics (AIMD) simulations. A large 4 × 3 supercell with sizes of 12.884 and 16.698 Å is adopted. By heating the Na$_2$BP specie to 300 K within a canonical ensemble, the MD simulation is last for 10 ps with a time step of 2.0 fs. At the end of the simulation, the final structure is carefully checked. The snapshots viewed from different directions are displayed in Figure S7a. Due to the thermal disturbance and electrostatic repulsive interactions, the borophosphene exhibits obviously corrugated distortion. But, no structural collapse occurs. After removing all Na atoms, the structural reversibility is verified by performing AIMD simulations to the residual corrugated borophosphene. After 10 ps simulation, the corrugated borophosphene can be fully recovered to its initial planar structure with slightly bending distortion (see Figure S7b), suggesting good cyclical stability of borophosphene.

In addition to the specific capacity, open circuit voltage (OCV) is another key factor to characterize the performance of SIBs, which can be deduced directly from the energy difference before and after Na intercalation. Considering that the charging/discharging processes of borophosphene-based anode can be expressed as:

$$\text{Na}_{x_1}\text{BP} + (x_2 - x_1)\text{Na}^+ + (x_2 - x_1)e^- \leftrightarrow \text{Na}_{x_2}\text{BP},$$

an average OCV can be derived according to the following formula:

$$\text{OCV} \approx \frac{E_{\text{Na}_{x_1}\text{BP}} + (x_2 - x_1)E_{\text{Na}} - E_{\text{Na}_{x_2}\text{BP}}}{x_2 - x_1},$$

where $E_{\text{Na}_{x_1}\text{BP}}$, $E_{\text{Na}_{x_2}\text{BP}}$, and $E_{\text{Na}}$ are the potential energies of $\text{Na}_{x_1}\text{BP}$, $\text{Na}_{x_2}\text{BP}$, and metallic Na bcc bulk, respectively. Figure 4 shows the average OCV of $\text{Na}_x\text{BP}$ as a functional of the Na content $x$ in the range of 0 to 0.5. In the low range ($0 \leq x \leq 0.083$), the OCV is predicted to be 0.838 V. Then, it drops down as the Na content x increases. Subsequently, a low voltage of 0.140 V is maintained when $x = 0.5$. For the whole range of 0 ~ 0.5, an average voltage profile is 0.367 V, which is comparable to 0.32 V of $\text{B}_2\text{S}$[29] and 0.41 V of $\text{PC}_3$.[46] Moreover, this average voltage is in the moderate range for the application in anode materials where the suitable operating voltage is reported to be lower than 1.5 V.[50] So, the borophosphene can be used as potential anode material for SIBs.

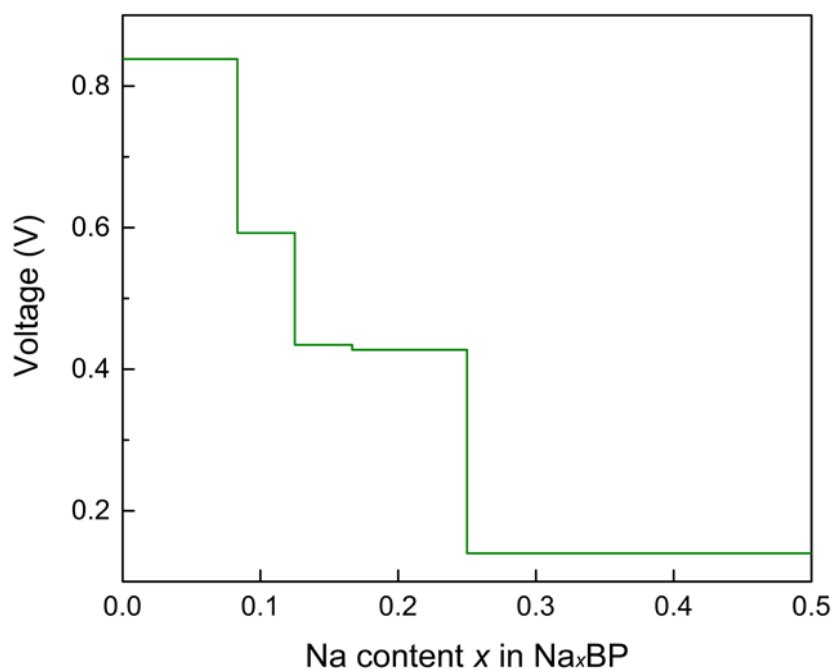

**Figure 4.** Average voltage as a functional of Na content $x$ in $\text{Na}_x\text{BP}$ species.

As we all know, the electrochemical performances of rechargeable batteries, including cyclability and rate capability are closely related to the electronic properties of electrode materials.[51] Therefore, we have calculated the electronic band structures of $Na_xBP$ species with $x$=0.08, 0.25 and 0.5. The results are shown in Figure 5. For comparison, the band structure of pristine borophosphene is also presented (see Figure 5a). Obviously, a Dirac cone can be observed between the $k$-points of $\Gamma$ and X. Thus, the pristine borophosphene is semi-metallic. Along the $k$-points of $\Gamma \rightarrow Y \rightarrow S$, both the highest valence band and the lowest conduction band are double degenerate. The degeneracy of these energy levels will disappear when the borophosphene absorbs Na adatoms with the content of $x$ = 0.08, as shown in Figure 5b. Because of the ionic interactions came from Na-B and Na-P bonds, the level of Dirac cone drops down 0.40 eV or so, and the borophosphene is metallic. For the rest $Na_xBP$ species with higher Na contents (see Figure 5c and 5d), their metallic properties suggest good electronic conductivity, which is advantageous for achieving high performance of anode material.

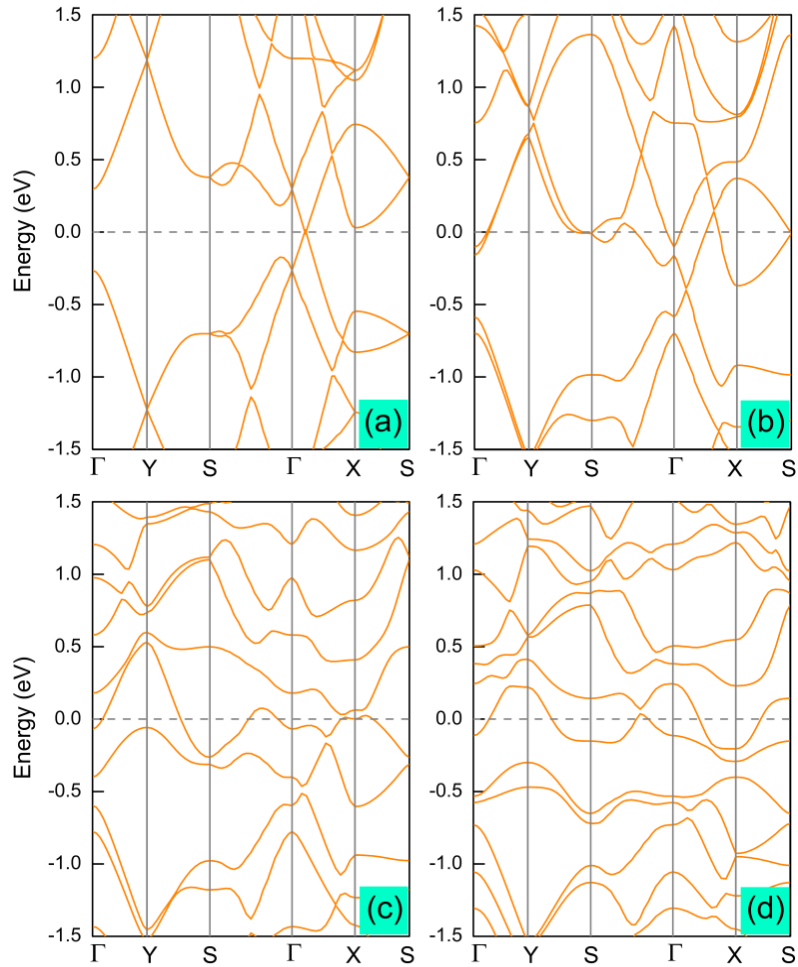

**Figure 5.** Electronic band structures of $Na_xBP$ species: (a) $x$=0.0 (i.e., pristine borophosphene), (b) $x$=0.08, (c) $x$=0.25, and (d) $x$=0.5. Dash line denotes the Fermi level.

In many 2D anode materials, the sodiation process can introduce an apparent expansion of surface volume, which can even result in the electrode fracture. To address this issue, we have performed fully structural optimization with cell size relaxing for all $Na_xBP$ species. The expansion of surface area can be estimated with the equation of $\varepsilon = (S-S_0)/S_0$, where $S$ and $S_0$ are the surface areas of sodiated and pristine borophosphene sheets, respectively. The corresponding results are shown in Figure S8. It is worthy to note that the $Na_xBP$ species with low Na contents exhibit tiny surface expansion (below 1 %). A maximum value no more than 2 % is obtained when $x = 1.0$. Such tiny surface expansion could help the borophosphene anode to maintain good cyclical stability during the charging and discharging processes. In our previous study, we have demonstrated that the borophosphene is very soft because of its small in-plane Young's modulus ranged from 126.2 to 145.8 N/m, which are much smaller than those of graphene (~ 340 ± 40 N/m),[52] $BC_3$ (~ 316 N/m),[53] and BN (~ 267 N/m).[54] These small modulus can result in favorable corrugated deformation rather than surface expansion for the sodiated borophosphene. Similar structure changes from a flat structure to a corrugated one can be observed in $B_2S$ monolayer.[29]

## 4. CONCLUSIONS

In summary, spin-polarized DFT calculations have been employed to systematically investigate the electrochemical performances of borophosphene as a potential Dirac anode material for SIBs. The results show that the borophosphene can spontaneously adsorb Na atoms with favorable binding strength, as well as a low diffusion energy barrier of 0.221 eV. A maximum stoichiometry of $Na_2BP$ can be achieved with a remarkable large theoretical specific capacity of 1282 mAh/g, which is one of the largest values reported in 2D anode materials for SIBs. AIMD simulations have revealed that the borophosphene exhibits good cyclability with no structural collapse during the charging/discharging process. Moreover, all the sodiated borophosphene present good electronic conductivity and tiny surface expansion. Our results above suggest that the borophosphene can be expected to use as a promising anode material with large specific capacity, high rate capability and good cyclical stability for SIBs.

**AUTHOR INFORMATION**

Corresponding Author

[*]E-mail: yzhang520@mail.xjtu.edu.cn (Y. Zhang).

**Notes**

The authors declare no competing financial interest.

## ACKNOWLEDGEMENTS

This work was supported by the Natural Science Fundamental Research Program of Shaanxi Province of China (Grant No. 2019JM-190), the National Natural Science Foundation of China (Grant No. 11774278 and 11774280), and Special Guidance Funds for the Construction of World-class Universities (Disciplines) and Characteristic Development in Central Universities.

of Monolayer Graphene. *Science* **2008**, *321*, 385-388.